\documentclass[12pt]{article}
\usepackage{geometry}  
\usepackage{amsmath}
\usepackage{amssymb}
\usepackage{hyperref}

\usepackage{graphicx,dcolumn,bm}
\usepackage[usenames,dvipsnames]{color}
\usepackage{mathrsfs}

\let\div\undefined
\DeclareMathOperator{\div}{div}

\DeclareMathOperator{\Tr}{Tr}

\begin{document}

\newcommand{\nino}[1]{\textcolor{BrickRed} {\bf#1}}
\newcommand{\detlef}[1]{\textcolor{Green} {#1}}
\newcommand{\shelly}[1]{#1}

\newcommand{\psicap}{\Psi}
\newcommand{\rr}{\mathrm{L}}
\newcommand{\rrr}{\rho}
\newcommand{\Om}{{\boldsymbol{\Omega}}}

\newcommand{\g}{\mathsf{g}}
\newcommand{\CS}{\mathcal{Q}}
\newcommand{\CSO}{\QS}

\newcommand{\bigcdot}{\boldsymbol{\cdot}}
\newcommand{\q}{{\boldsymbol{q}}}
\newcommand{\p}{{\boldsymbol{p}}}
\newcommand{\kk}{{\boldsymbol{k}}}
\newcommand{\vv}{{\boldsymbol{v}}}

\newcommand{\C}{C}

\newcommand{\ve}{\boldsymbol{e}}

\newcommand{\qs}{q}

\newcommand{\QA}{\pmb{\mathcal{Q}}}
\newcommand{\QS}{\mathcal{Q}}

\newcommand{\bomega}{\boldsymbol{\omega}}
\newcommand{\Qs}{Q}

\newcommand{\X}{\boldsymbol{\mathsf{X}}}
\newcommand{\Y}{\boldsymbol{\mathsf{Y}}}

\newcommand{\nablab}{\boldsymbol{\nabla}}
\newcommand{\Q}{\boldsymbol{Q}}
\newcommand{\spot}{\mathscr{V}}

\newcommand{\new}[1]{\textcolor{NavyBlue} { #1}}

\newcommand{\vect}[1]{\vec{\mathsf{#1}}}

\newcommand{\Hab}{\mathsf{H}}
\newcommand{\bHab}{\doublehat{\Hab}}
\newcommand{\bHabbb}{\triplehat{\Hab}}
\newcommand{\bHabb}{\mathcal{H}}
\newcommand{\bHabh}{\widehat{\mathsf{H}}}

\newcommand{\hHab}{\widehat{\mathsf{H}}}
\newcommand{\Op}{\mathsf{L}}
\newcommand{\euDelta}{\mathsf{\Delta}}

\newcommand{\lDeltag}{\widehat{\Delta}_B}

\newcommand{\Sc}{Schr\"odinger}
\newcommand{\psibh}{\widehat{\Psi}}
\newcommand{\psib}{\doublehat{\Psi}}
\newcommand{\psibb}{\triplehat{\Psi}}
\newcommand{\Psiq}{\psi}
\newcommand{\hPsiq}{\widehat{\psi}}
\newcommand{\psiq}{\psi}
\newcommand{\F}{$\mathscr{F}$ }
\newcommand{\E}{$\mathscr{E}$ }
\newcommand{\muF}{$\mu_\mathscr{F}$ }
\newcommand{\z}[1]{\textbf{#1}}

\newcommand{\ovq}{\vect{\mathcal{q}}}
\newcommand{\ovqc}{\widetilde{\boldsymbol{{q}}}}

\newcommand{\lift}[1] {\boldsymbol{#1}}
\newcommand{\liftw}[1] {\widehat{#1}}

\newcommand{\jac}{\mathfrak{J}}

\newcommand{\whpsi}{\widehat{\; \Psi\;}}
\newcommand{\whpsic}{\widehat{\; \psi\;}}

\newcommand{\gaupsi}{\widehat{\Psi}}

\newcommand{\gauconpsi}{\widehat{\psi}}

\newcommand{\zerogaupsi}{\widehat{\Psi}_{S}}
\newcommand{\firstgaupsi}{\widehat{\Psi}_{1}}

\newcommand{\secondgaupsi}{\widehat{\Psi}_{2}}
\newcommand{\thirdgaupsi}{\widehat{\Psi}_{3}}

\newcommand{\zerogauH}{\bHabh_{S}}

\newcommand{\firstgauH}{\bHabh_{1}}
\newcommand{\secondgauH}{\bHabh_{2}}
\newcommand{\thirdgauH}{\bHabh_{3}}

\newcommand{\Ppath}{\mathbb{P}}
\newcommand{\Epath}{\mathbb{E}}

\newcommand{\Pconf}{{P}}

\renewcommand{\psibh}{\firstgaupsi}
 \renewcommand{\bHabbb}{\thirdgauH}
 \renewcommand{\psibb}{\thirdgaupsi}
\renewcommand{\bHabb}{\zerogauH} 
 
 \renewcommand{\psib}{\secondgaupsi}
 \renewcommand{\bHab}{\secondgauH}

\title{ \shelly{Remarks About the Relationship Between Relational Physics  and a Large Kantian Component of the Laws of Nature}}

\author{
Sheldon Goldstein\footnote{Departments of Mathematics, Physics and
     Philosophy, Hill Center, Rutgers, The State University of New  
     Jersey, 110 Frelinghuysen Road, Piscataway, NJ 08854-8019, USA.
     E-mail: oldstein@math.rutgers.edu} 
     and Nino Zangh\`\i \footnote{Dipartimento di Fisica dell'Universit\`a
     di Genova and INFN sezione di Genova, Via Dodecaneso 33, 16146
     Genova, Italy. E-mail: zanghi@ge.infn.it}
}

\date{\today}

\maketitle

\begin{abstract}
Relational mechanics is a reformulation of mechanics (classical or quantum) for which space is relational. This means that  the configuration of an $N$-particle system is  a shape,  which is what remains when the effects of rotations, translations and dilations are quotiented out. This reformulation of mechanics  naturally leads to a relational notion of time as well, in which a history of the universe is just a curve in shape space without any reference to a special parametrization of the curve given by an absolute Newtonian time.  When relational mechanics  (classical or quantum) is regarded as fundamental, the usual descriptions in terms of absolute space and absolute time emerge merely as corresponding to the choice of a gauge. This gauge freedom  forces us to recognize that what we have traditionally regarded as fundamental in physics might in fact be imposed by us through our choice of gauge. It thus imparts a somewhat Kantian aspect to physical theory.\end{abstract}

\tableofcontents

\section{Relational Physics}\label{sec:rph}
We shall address here  three basic questions concerning relational physics: What is it? What is its point? What  can we \shelly{learn} from it? These issues have been addressed in a long paper \cite{QMSS} published in 2020 with our friend and colleague Detlef \shelly{D\"urr,} who   passed away in January 2021. 

Relational physics has been revived in relatively recent times by Julian Barbour and Bruno Bertotti, who, in a very
inspiring and influential paper published  at the beginning of the eighties \cite{Barbour:1982gha}\shelly{,
transformed} a long standing philosophical controversy
about  the nature of space and time into a well-defined physical problem. Our  main focus here will be on gauge freedom and objectivity \shelly{versus} subjectivity, probability, and the non-normalizability of the quantum state.

\subsection{Relational Space}\label{sec:rs}
The philosophical  issue about the nature of space dates back to the  dispute between Isaac Newton, who {favored} {and argued for the need of } an absolute theory of space and time,  and Gottfried Wilhelm Leibniz, who insisted upon a relational approach,  also defended by Ernst Mach in the 19th  century.  \shelly{To} put it briefly,
the gist of the relational approach is that overall position, orientation, and size are not relevant.
This can be  clarified by means of a very elementary and simplified  model of the universe.

Suppose we are given the configuration of a universe of $N$  particles.
And suppose we translate every particle of the configuration in the same 
direction by the same
amount. From a physical point of view it seems rather natural to
take the relational \shelly{perspective} that the two
configurations of the universe so obtained are physically equivalent or
identical. Similarly for any rotation. Going one step further, one regards two
configurations of the universe differing  only  by a dilation, i.e. by a uniform
expansion or contraction, as representing  in fact the same physical state of
the universe. 
The   space of all genuinely  physically different possible
 configurations so obtained---taking into account translations, rotations,  and
dilations---is  usually called {\it shape space}.  
The name shape space is  indeed natural: only the shape of a configuration of
particles is relevant, not its position or orientation  or overall size.

Shape space should be contrasted with {\it absolute configuration space},
the totality of configurations 
of $N$ points in Euclidean three-dimensional space.

\subsection{Relational Time}

Once shape space is taken as fundamental, one should address how shapes evolve. While  positing  an  absolute Newtonian  time as in classical mechanics is compatable with relational space, it is  natural to consider  a relational notion of  time as well. According to relational time,  global changes of speed of the history of the universe give physically equivalent representations, so that a history of the universe is just a curve in shape space without any reference to a special parametrization of the curve  given by \shelly{any sort of} absolute  Newtonian  time. 

Relational time should be contrasted with absolute or {\em metrical time}.
By metrical time we refer to any objective physical coordination of the configurations along a  geometrical path  in a configuration space with the points of a one-dimensional continuum: a (continuous) mapping from the continuum onto the path. The continuum is usually represented by the real numbers, but it need not be. However, it should be physically distinct from the particular continuum that is the path itself.

Understood in this way, metrical time does not exist, from the relational point of view, for the universe as a whole. However, for subsystems of the universe metrical time naturally emerges: the continuum with which the geometrical path corresponding to the evolution of the subsystem is coordinated can be taken to be the path of its environment, with the obvious mapping between the paths.

\subsection{Relational versus Relativistic}

Relational physics should be contrasted with relativistic physics. A simple point: In relational physics  the traditional separation of space and time is retained. While configuration space is replaced by shape space, and time becomes non-metrical,  shape space  retains an identity separate and distinct from that of (non-metrical)  time. This is in obvious contrast with relativistic physics, in which space and time lose their separate identities and are merged into a space-time.

Perhaps the most characteristic feature of relativity is the absence of absolute simultaneity. Not so for relational physics. Since it retains the separation of space and time,  an absolute simultaneity is built into the very structure of relational physics as described here. Nonetheless, there is a sense in which simultaneity is lost.
With relational time the notion of the shape of the universe at ``time $t$'' is not physically meaningful. And with what is meaningful---geometrical paths in the space of possible   shapes---one can no longer meaningfully compare or ask about the configurations for two different possible histories at the same time. Given the actual configuration of the universe, it is not meaningful to ask about the configuration of an alternative history at that time without further specification of exactly what that should mean.

One may inquire \shelly{whether} the relational point of view can be merged with or extended to relativity. Can we achieve a  relational understanding of space-time? General relativity is certainly a step in that direction, but it does not get us there. Space-time in general relativity is metrical---in a way that neither space nor time are in relational physics. A complete extension, if at all possible, is a real challenge.

Another possibility is that relativity is not fundamental, but is, instead,  a consequence of a suitable choice of gauge. This possibility, which is suggested by the work of Bryce DeWitt \cite{dewitt1970spacetime} and Barbour and {coworkers}  \cite{barbour2002relativity} would be worth carefully exploring. For a bit  of elaboration on this, see Section \ref{sec:geo}.

\subsection{Relational Space, Relational Time, and Gauge Freedom}
When physical theories  are formulated in shape space, one should consider first the simplest ones, namely the ``free'' theories  based only on the geometrical structures provided by the metric, without invoking any potential. This is in contrast with theories formulated in absolute space, for which free theories can't begin to account for the experimental data.  It is then natural to ask: when we represent the theories in absolute space, what form do the laws of motion  take? Is the representation unique or are there various representations yielding \shelly{different-looking }laws of motion, some  unfamiliar and some more or less familiar? Moreover, do  interacting theories emerge with nontrivial interactions, although in shape space the motion is free? 

To  answer  these questions it is helpful to represent absolute configuration space in geometrical terms as a fiber bundle, with shape space as base manifold and the fibers generated by the similarity group, i.e, by translations, rotations and dilations, which acting on configurations \shelly{yield,}  from a relational point of view,  physically equivalent states.  A representation in absolute configuration space of the motion in shape space is then given by  a ``lift'' of the motion from the base into  the fibers (more details below). 

Such lifts can rightly be called gauges. In the classical case it turns out that in some gauges the law looks unfamiliar but there is (at least) one gauge in which,  after performing a time change (representing indeed another gauge freedom when also time is seen as relational),  the law of motion is Newtonian with a potential appearing. 
So, it turns out that \shelly{relational space and relational time form  a harmonious combination.} More or less the same is true for the quantum case. And similarly, as hinted above, the same might be true for general relativity.

\subsection{Why \shelly{Go} Relational?}

One reason  for turning to relational physics is this. \shelly{Only relational distinctions concerning space and time are observable. Thus, so the argument might continue,} since the positions of bodies are individuated only \shelly{by}  relative distances and angles, from an observational point of view, only shapes or \shelly{relational} times are meaningful, while absolute space, absolute time, and even absolute spacetime, do not have any clear observational meaning. 
While having some force,  this \shelly{argument} is too positivistic to be taken as the main reason to go relational.

Here is another reason. Since the long-standing philosophical controversy
about  the nature of space and time    has now become a well-defined problem in theoretical physics,  we have a professional reason to 
go relational. \shelly{It seems worth exploring how far one can go in formulating physics in minimal relational terms.}

A stronger reason, which is  indeed the  one we favor, has to do with 
the status and the nature of the wave function. Since at this stage, this sounds totally incomprehensible, 
we shift gears and  turn to  quantum mechanics.

\section{Quantum Mechanics}
Standard quantum mechanics is problematic. It is often said  that the fundamental problem with quantum mechanics is the {\it  measurement problem}, or, more or less equivalently, Schr\"odinger's cat paradox. However,  the measurement problem, as important as it is, is nonetheless but a manifestation of a more basic difficulty with standard quantum mechanics: it is not at all clear what quantum theory is about. Indeed, it is not at all clear what quantum theory actually says. Is quantum mechanics fundamentally about measurement and observation?  Or is it about the behavior of suitable fundamental microscopic entities, elementary particles and/or fields?
To use the words of Bell, standard formulation of quantum mechanics are  ``unprofessionally vague and ambiguous.'' A formulation free of these shortcomings is 
Bohmian mechanics \cite{bohm1952suggested,bell1987speakable,durr1992quantum,durr2012quantum,bricmont2016}.  

\subsection{Bohmian Mechanics}
Bohmian mechanics  is a theory
providing a description of reality, compatible with all of the quantum
formalism, but free of any reference to observables or observers \shelly{in its formulation}. In Bohmian mechanics a system of
particles is described in part by its wave function, evolving according to
Schr\"odinger's equation, the central equation of quantum theory. However,
the wave function provides only a partial description of the system. This
description is completed by the specification of the actual positions of
the particles. The latter evolve according to the ``guiding equation,''
which expresses the velocities of the particles in terms of the wave
function. Thus in Bohmian mechanics the configuration of a system of
particles evolves via a deterministic motion choreographed by the wave
function.

Bohmian mechanics might be regarded as  the minimal completion of Schr\"odinger's equation,
for a non-relativistic system of particles, to a theory describing a genuine
motion of particles. For Bohmian mechanics the state of a system of $N$
particles is described by its wave function $\psicap{} = \psicap{}(\vect{q}_1,
\dots ,\vect{q}_N) = \psicap{}(\q )$, a complex- (or spinor-) valued function on
the space of possible configurations $\q $ of the system, together with its
actual configuration $\Q$ defined by the actual positions $\vect{Q}_1,
\dots ,\vect{Q}_N$ of its particles. The theory is then defined by two
evolution laws. One is {\it Schr\"odinger's equation}
\begin{equation}
\label{Schroedinger}
  i\hbar \frac{\partial \psicap{}}{\partial t} = H\psicap{}\,,
\end{equation}
for $\psicap{}=\psicap{}_t$, the wave function at time $t$, where $H$  is the
non-relativistic (Schr\"odinger) Hamiltonian, containing the masses \shelly{$m_{\alpha}$, $\alpha=1, \ldots, N$,} of the
particles and a potential energy term $V$. For spinless particles, it is of the form
\begin{equation}
H=-\sum_{\alpha=1}^{N}
\frac{{\hbar}^{2}}{2m_{\alpha}}\vect{\nabla}^{2}_{\alpha} + V\,,
\label{sh}
\end{equation}
where $\vect{\nabla}_\alpha = \frac{\partial\,\;}{\partial \vect{q}_\alpha}$ is the gradient with respect to the position  of the $\alpha$-th particle.
The other law is the {\it the guiding law},  given by the equation
\begin{equation}\label{Bohm}\
  \frac{d\vect{Q}_\alpha}{dt} = \frac{\hbar}{m_\alpha} \mathrm{Im} \frac{\psicap{}^*
  \vect{\nabla}_\alpha\psicap{}}{\psicap{}^* \psicap{}} ( \vect{Q}_1, \dots ,\vect{Q}_N )\,.
\end{equation} for $\Q=\Q(t)= ( \vect{Q}_1(t), \dots ,\vect{Q}_N (t) )$, the configuration at time $t$ \shelly{(with $\mathrm{Im}$ denoting the imaginary part of the complex quantity that follows it).} If $\psicap{}$ is
spinor-valued, the products in numerator and denominator in \eqref{Bohm} should be
understood as scalar products. If external magnetic fields are present, the
gradient should be understood as the covariant derivative, involving the
vector potential.  For spinless particles, setting \shelly{$\psicap{} = R e^{iS/\hbar}$ (the polar representation of $\psicap$)}, the guiding equation \eqref{Bohm} becomes
\begin{equation}
 \frac{d\vect{Q}_\alpha}{dt} = \frac{\hbar}{m_\alpha} \mathrm{ Im } \frac{  \vect{\nabla}_\alpha \psicap{}}{ \psicap{}} ( \vect{Q}_1, \dots ,\vect{Q}_N )
 = \frac{1}{m_\alpha} \vect{\nabla}_\alpha S ( \vect{Q}_1, \dots ,\vect{Q}_N )
 \label{gl}
\end{equation}

For an $N$-particle system Schr\"odinger's equation and the guiding equation, together
with the detailed specification of the Hamiltonian $H$,  completely define the
Bohmian motion of the system.

\subsection{OOEOW}
Bohmian mechanics  is for us   the simplest version of quantum mechanics\shelly{: Given Schr\"odinger's equation,  it has  the Obvious Ontology Evolving in the Obvious Way.} 

For a theory of {\em particles}, in addition to the wave function you've got the actual positions of the particles. This is a rather obvious ontology.
We usually emphasize this point by saying that the positions of the particles provide the {\em primitive ontology} of the theory. In so saying we wish to convey that the whole point of the theory---and the whole point of the wave function---is to define a motion for the particles, and  it's in terms of this motion that pointers end up pointing and experiments end up having results, the kinds of  results that it was the whole point of quantum mechanics to explain. So the connection to physical reality in the theory is  via what we're calling the primitive ontology of the theory, in Bohmian mechanics the positions of the particles.

The only thing Bohmian mechanics adds is a first-order equation of motion for how the positions of particles evolve. This equation, the new equation in Bohmian mechanics, is kind of an obvious equation. It is more or less the first thing you would guess if you asked yourself, What is the simplest motion of the particles that could reasonably be defined in terms of the wave function?

There are indeed many  (more or less) obvious ways of guessing the velocity field  function $\vv=\vv^{\psicap}$ in the right hand side of \eqref{Bohm}. The simplest way, for particles without spin,  is the following:  Begin with the de Broglie relation ${\p} = \hbar {\kk}$, a remarkable and mysterious distillation of the experimental facts
associated with the beginnings of quantum theory---and itself a relativistic reflection of the first quantum equation, namely the Planck relation $E= h{\it \nu}$.  The de Broglie relation  connects
a particle property, the momentum ${\p}=m{\vv}$, with a wave
property, the wave vector $ {\kk}$. Understood most simply, it says that the velocity of a
particle should be the ratio of $\hbar {\kk}$ to the mass of the
particle. But the wave vector ${\kk}$ is defined  only for a plane
wave. For a general wave $\psi$, the obvious generalization of $ {\kk}$
is the local wave vector $\mathbf{\nabla}S({\q})/\hbar$, where $S$ is the phase of the wave function (defined by its polar representation, see \shelly{above}).
With this
choice the de Broglie relation becomes  $\vv= \nabla
S / m$, the right hand side of which is our first guess for $\vv^{\psi}.$

We note that the  de Broglie relation also immediately yields Schr\"odinger's
equation, giving the time evolution for $\psi$, as the simplest wave
equation that reflects this relationship. This is completely standard. In this simple way, the defining equations of Bohmian mechanics can be regarded as flowing in a natural manner from the first quantum equation $E= h\nu$.

 The quantum continuity equation is the key 
{for a route which is meaningful  also for particles with spin}. This equation, an immediate consequence of 
Schr\"odinger's equation, involves a quantum probability density $\rho$ and a quantum probability current $J$. Since densities and currents are classically related by $J= \rho \mbox{\sl v}$, it requires little imagination to set $\vv^\psi= J/\rho$.

 Another way to arrive at a formula for $\vv^{\psi}$ is to invoke symmetry. Since the space-time symmetry of the non-relativistic  Schr\"odinger equation is that of rotations, translations, time-reversal, and invariance under Galilean boosts, it is natural to demand that this Galilean symmetry be retained when Schr\"odinger's equation is combined with the guiding equation. As described in \cite{durr1992quantum}, this leads to a specific  formula for $\vv^{\psi}$ as the simplest possibility.

There are  several {other} natural ways to arrive at $\vv^{\psi}$, but we shall give no more. All these routes, those we've explicitly mentioned and those to which  we've alluded, yield in fact exactly the same formula for $\vv^{\psi}$, though the explicit form may appear different in some cases.

\subsection{The Implications  of Bohmian Mechanics}
While the {\it formulation\/}  of Bohmian mechanics does not involve the notion of
quantum observables, as given by self-adjoint operators---so that its
relationship to the quantum formalism \shelly{(the familiar axioms of quantum theory)} may at first appear somewhat obscure---it can in
fact be shown that  Bohmian mechanics  embodies
quantum randomness, as expressed by Born's rule, and \shelly{the entire}
quantum formalism  based on self-adjoint operators as \shelly{observables} as the very expression of its empirical import
\cite[Ch.2 and 3]{durr2012quantum}.  The most notably implication of Bohmian mechanics is the fact that
in a universe governed by Bohmian mechanics there are sharp, precise, and irreducible limitations on the possibility of obtaining knowledge, limitations that can in no way be diminished through technological progress leading to better means of measurement. This {\em absolute uncertainty} is in precise agreement with Heisenberg's uncertainty principle.

We shall review  below the key points that lead to these implications.

 \subsubsection{The Conditional Wave Function}

In physics we are usually concerned not with the entire universe but with subsystems of the universe, for example with a hydrogen atom or a pair of entangled photons. The quantum mechanical treatment of such systems  involves the quantum state of that system, often given by its wave function---not the wave function of the universe. 
Yet, the notion of the wave function of a subsystem is  rather elusive. In standard quantum mechanics  the state of a subsystem is usually described in terms of its  reduced density matrix.

However,
Bohmian mechanics provides a precise formulation and understanding of this notion in terms of the conditional wave function \cite{durr1992quantum}
\begin{equation}\label{conprobfor}
\psi(x)=\Psi(x,Y)\,,
\end{equation}
where $\Psi=\Psi(q)=\Psi(x,y)$ is the wave function of the universe, 
with $x$ and $ y$   the generic variables for the configurations of the system and its environment  \shelly{(everything else)}, respectively, and  where $Y$ is  the actual configuration of the environment. The conditional wave function of a Bohmian system behaves exactly as one would expect the wave function of a system to behave, with respect to both  dynamics and statistics.

 \subsubsection{The Conditional Probability Formula}
In Bohmian mechanics, for a non-relativistic system of particles, the configuration of a system is regarded as random, with randomness corresponding to the
 quantum equilibrium distribution $\mu^\psi$ given by $|\psi|^2dq$. What this actually means, in a deterministic theory such as Bohmian mechanics, is a delicate matter, involving a long story \cite{durr1992quantum} with details and distinctions that we shall ignore here. However, a crucial ingredient for that analysis---for an understanding of the origin of quantum randomness in a universe governed by Bohmian mechanics---is the {\it fundamental conditional probability formula} for the conditional distribution of the configuration $X_t$ of a system at time $t$ given that of its environment~$Y_t$ at that time:
\begin{equation}\label{condprob}
 P^{\Psi_0}(X_t\in dx\, | Y_t) = |\psi_t(x)|^2 dx, 
\end{equation}
 where $\Psi_0$ is the initial wave function of the universe and $P^{\Psi_0}$ is the probability distribution on trajectories arising from the Bohmian dynamics with an initial quantum equilibrium distribution, and $\psi_t$ is the (normalized) conditional wave function \eqref{conprobfor} of the system at time $t$.
 
 A crucial ingredient  in proving \eqref{condprob} is {\em equivariance} \cite{durr1992quantum}:  if  at any time   the  \shelly{configuration of a Bohmian system} is randomly distributed 
 according to \shelly{$|\Psi|^2$  at that time} then at any other time $t$ it will be distributed according to 
 $|\Psi_t|^2$. Equivariance  is an immediate consequence of the continuity equation  arising from the Schr\"odinger equation:
 \begin{equation}  \label{continuityequation}
 \frac{\partial \rho^\Psi}{\partial t} + \div J^\Psi = 0\,,
 \end{equation}
with   $\rho^\Psi= |\Psi|^2$, the quantum equilibrium distribution,  and 
\begin{equation} \label{quantumcurrent}
J^\Psi = \rho^\Psi \vv^\Psi \,,
\end{equation}
 the quantum probability current,  where $\vv^\Psi$ is  the Bohmian velocity
in the right hand side of \eqref{gl}.

\subsection{Bohmian Motion on a Riemannian Manifold}

We conclude this section with a mathematical observation, whose importance will become clear later on when we \shelly{discuss} the quantum mechanics of shapes.
Note   that, given $V$,  the Bohmian mechanics defined by equations  \eqref{Schroedinger}, \eqref{sh}, and \eqref{gl} depends only upon the Riemannian structure $g=g_e$ given by the standard Euclidean metric on configuration space
$[g_e]_{ij} = m_{\alpha_i}  \delta_{ij}$,  where the $i$-th component 
 refers to the $\alpha_i$-th particle.  In
terms of this Riemannian structure, the evolution equations \eqref{gl} and  \eqref{sh} 
 become
\begin{align}\label{eq:bmonmany1x}
\frac{dQ}{dt} &= \hbar  \, \mathrm{ Im } \frac{\nabla_g \psicap{}}{\psicap{}} \\
\label{eq:bmonmany2x}
i\hbar \frac{\partial \psicap{}}{\partial t} &= -\frac{\hbar^2}{2} \Delta_g \psicap{} + V\psicap{}\,,
\end{align}
where  $\Delta_g$ and  $\nabla_g$ are,
respectively, the Laplace-Beltrami operator and the gradient on the
configuration space equipped with this Riemannian structure. But there is nothing special about 
this particular Riemannian structure. Indeed, equations \eqref{eq:bmonmany1x} and 
 \eqref{eq:bmonmany2x}  as such hold  very generally  on {\em any} Riemannian manifold. 
 Thus, the formulation of a Bohmian dynamics on a Riemannian manifold requires only  as  basic 
ingredients  the differentiable and metric structure of the
manifold.

\section{Quantum Puzzles}\label{sec:qp}
%

The Bohmian perspective allows \shelly{us to illuminate}  some quantum puzzles, somewhat hidden in other formulations of quantum mechanics. \shelly{These puzzles find their natural resolution in a proper relational reformulation of quantum mechanics. We shall elaborate in later sections.}

\subsection{The Wave Function of the Universe and Entropy}\label{sec:wfe}
The fundamental equation for the wave function of the universe in canonical quantum cosmology is the Wheeler-DeWitt equation 
\cite{dewitt1967}
\begin{equation} \mathscr{H}\varPsi=0\,, \label{wdw} \end{equation}
for a wave function $\Psi(q)$ of the universe, where  $q$ refers to 3-geometries and to whatever other stuff is involved, all of which correspond to structures on a 3-dimensional space.  In this equation $\mathscr{H}$ is a sort of generalized Laplacian, a cosmological version of a \Sc\ Hamiltonian $H$. And like a typical $H$, it involves nothing like an explicit time-dependence. But unlike \Sc's equation, the  Wheeler-DeWitt equation has on one side, instead of a time derivative of $\Psi$, simply 0. Its natural solutions are thus \shelly{time-independent.}  

That this is so is in fact the {\em problem of time} in quantum cosmology. We live in a world where things change. But if the basic object in the world is a timeless wave function, how does change come about? Much has been written about this problem of time.  A great many answers have been proposed. But what we want to emphasize here is that from a Bohmian perspective the timelessness of  $\Psi$ is not a problem. Rather it is just what the doctor ordered. 

The fundamental role of the wave function in Bohmian mechanics is to govern the motion of something else. Change fundamentally occurs in Bohmian mechanics not so much  because the wave function changes but because the thing $Q$  \shelly{that it governs changes,} according to a law 
\begin{equation}\label{ge1}
dQ/dt=v^{\varPsi}(Q)
\end{equation}
determined by the wave function. The problem of time vanishes entirely from a Bohmian point of view. 

But there is a difficulty that arises in connection with the 
stationarity of the universal wave 
function. 
According to standard formulations of quantum statistical mechanics, thermodynamics seems to depend only upon the wave function. But if the initial wave function is 
stationary, it can't be responsible for the irreversible behavior of 
our universe, \shelly{and in particular of entropy increase.} So what is the origin of the arrow of time? \shelly{And why does entropy} 
increase?

\subsection{Nonnormalizability of the Wave Function of the Universe}
The solutions of Wheeler-DeWitt equation typically fail to be normalizable.
So what could the associated non-normalizable  ``probabilities" given by $|\Psi|^2 $ physically mean? How \shelly{can} the quantum equilibrium analysis we hinted \shelly{at above, that allows  us to recover the entire quantum formalism,  get off the ground for such a non-normalizable measure}?

\subsection{Why vs. What}\label{sec:wvsw}
From a Bohmian perspective there is another puzzle. One may ask,   Why \shelly{should the motion of  configurations   be} governed by  a wave function?  

This problem   is rather subtle and elusive, hardly understandable at all in the standard formulation of quantum mechanics, as well as in other more precise versions of quantum mechanics, such as Many Worlds or GRW, in which the wave function is all there is.  We believe that  relational physics  sheds light on this problem and shall elaborate on this in Section \ref{sec:RQM}. \shelly{We also wish to note, concerning quantum mechanics, the contrast between asking why and asking what. One often finds derivations of quantum mechanics from some principle or other, which thus seem to answer the why question. But in most, if not all, such cases, the answer leaves us with little or no understaning of what quantum mechanics actually says about physical reality. This is not good. The appropriate time to ask why is after the what question has been satisfactorily addressed. With Bohmian mechanics we are at such a time.}

\section{Wave Function as Law}\label{sec:wfal}

The most puzzling issue in the foundations of quantum mechanics is perhaps that of the status of the wave function of a system in a quantum universe.
We have suggested elsewhere that  the wave function should be regarded as nomological, nomic---that it's really more in the nature of a law than a concrete physical reality. 

Thoughts in this direction might \shelly{arise} when you consider the unusual \shelly{way Bohmian mechanics is formulated, and the unusual way the wave function behaves in Bohmian mechanics.} The wave function of course affects the behavior of the configuration, i.e., of the particles. This is expressed by the guiding equation \eqref{ge1}. But in Bohmian mechanics there's no back action, no effect in the other direction, of the configuration upon the wave function, which evolves autonomously via \Sc's equation, in which the actual configuration $Q$ does not appear. 

A second point is that for a multi-particle system the wave function $\psicap{}(\q)=\psicap{}(\vect{q}_1, \dots,  \vect{q}_N)$ is not  a \shelly{field on physical space, but on configuration space,} the set of all hypothetical configurations of the  system. For a system of more than one particle that space is not physical space. What  \shelly{this} suggests to us is that you should think of the wave function as describing a law and not as some sort of concrete physical reality. After all \eqref{ge1} is an equation of motion, a law of motion, and the whole point of the wave function here is to provide us with the law, i.e., with the right hand side of this equation.

\subsection{$\psi$ vs. $\Psi$}
There are, however, problems with regarding the wave function as nomological. 
Laws aren't supposed to be dynamical objects, they aren't supposed to change with time, but the wave function of a system typically \shelly{does}. And laws are not supposed to be things that we can control. But the wave function is often an initial condition for a quantum  system. We often, in fact, prepare a system in a certain quantum state, that is, with a certain wave function. We can in this sense control the wave function of a system. But we don't control a law of nature. This makes it a bit difficult to regard the wave function as nomological.

But with regard to this difficulty it's important to recognize, as already \shelly{suggested},  that there's only one wave function we should be worrying about, the fundamental one, the wave function $\Psi$ of the universe. In Bohmian mechanics, the wave function $\psi$ of a subsystem of the universe is defined in terms of the universal wave function $\Psi$ according to \eqref{conprobfor}. Thus, to the extent that we can grasp the nature of the universal wave function, we should understand as well, by direct analysis, also the nature of the objects that are defined in terms of it, and in particular we should have no further fundamental question about the nature of the wave function of a subsystem of the universe.


Note that from this perspective, the most important equation in quantum mechanics, the time-dependent Schr\"odinger's equation should be regarded as a phenomenological equation describing how the conditional wave function $\psi$  of a subsystem 
evolves---under suitable conditions of decoupling from its environment---\shelly{in a universe having a suitable stationary universal  wave function $\Psi$.}


\subsection{\shelly{A Kantian Component  of Laws of Nature}}

 Laws are supposed to be simple. But here a distinction should be made. Consider the velocity field in the right hand side of \eqref{ge1},
$$\shelly{v^\Psi\!(Q)} = v (\Psi, Q)  \,.$$
If we regard  $\Psi$  as a law, $v^\Psi $ should be a simple function of $Q$. 
On the other hand, if we regard  $\Psi$  as {\em part } of a law, $v (\Psi, Q)$ should be a simple function of $\Psi$ and $Q$.

\shelly{So we would like to find a compelling law of motion, or a compelling principle yielding such a law, that can be expressed in terms of a $v^\Psi$ for a suitable $\Psi$. We don't have that. But, as we shall explain, the flexibility and gauge freedom afforded by relational physics might help us reach our goal.}

\shelly{Ideally, the suitable $\Psi$ referred to above should be simple. Failing that, it should at least obey a simple law, such as the Wheeler-DeWitt   equation.
From} a fundamental point of view, it might be a complete accident that $\Psi$ obeys such an equation. It might just happen to do so. The fact that the equation is satisfied might have nothing to do with why the fundamental  dynamics is of the form \eqref{ge1}. But as long as $\Psi$ does satisfy the equation, by accident or not, all the consequences of satisfying it \shelly{would }follow.

The point here is to distinguish what is really fundamental  from what we have  traditionally regarded as fundamental. The latter \shelly{might}, in fact, be imposed by us. We express this  by saying that the laws of nature \shelly{might have a Kantian component.  And that Kantian component might be much larger than suggested so far, as will} become clearer in the next section devoted to relational quantum mechanics.

\section{Relational Quantum Mechanics}\label{sec:RQM}
 At the end of Section \ref{sec:rph} we said that 
the stronger reason for us to go relational  has to do with understanding 
the status and the nature of the wave function. At the beginning of Section \ref{sec:qp},
we claimed that the quantum puzzles presented  there find their natural resolution in a proper relational reformulation of quantum mechanics. 

It's time to  intertwine all these threads. And the key question to start with is the one we  asked in the last paragraph of the previous  section, What is really fundamental?

\subsection{Shape Space as  \shelly{the} Fundamental Configuration Space}\label{sec: spafcs}
Let us go back to the toy model of \shelly{an $N$-particle universe}  introduced in Section \ref{sec:rs}. The 
space of all \shelly{its genuinely  physically }different possible
 configurations  is {\it shape space}.  Mathematically, \shelly{this} is described as follows.

 \subsubsection{Shape Space}
The totality of configurations $\q= (\vect{q}_1, \ldots , \vect{q}_N)$  of $N$ points  in Euclidean three-dimensional space 
forms the  configuration space $\QA= \{ \q\} =\mathbb{R}^{3N}$ of an $N$-particle system. We shall call 
$\QA$ the {\em absolute configuration space}.
On $\QA$ \shelly{there is a natural  action of}   the similarity transformations of Euclidean space, namely rotations, translations and dilations, since each of them acts naturally on each component of the configuration vector \shelly{in  physical Euclidean space.}
The totality of such transformations form the group $G$ of 
{\em similarity transformations} of Euclidean space.  Since the shape of a configuration is \shelly{what remains} when the effects associated with  rotations, translations and dilations are filtered away,
the totality of shapes, i.e.,   the {\em shape space}, is  the quotient space $\QS\equiv \QA/G$,  the set of equivalence classes
with respect to the  equivalence relations provided  by the similarity transformations of Euclidean space.

As such, shape space is not in general a manifold. To transform it into a manifold some massaging is needed (e.g., by excluding from $\QA$ coincidence points and collinear configurations), but we shall  not enter into this.\footnote{For more details on this issue, see, e.g., \cite{le1993riemannian} and \shelly{the references} therein.} Here we shall assume that the appropriate massaging of  $\QA$ has been  performed  and that  $\QS$ is a manifold. Since the group of similarity transformations has dimension $7$ (3  for rotations + 3 for translations + 1 for dilations), the dimension of $\QS\equiv \QA/G$ is $3N- 7$ \shelly{for $N\geq 3$}. For $N=1$ and $N=2$  shape space is trivial (it contains just a single point).  $N=3$  corresponds to the  simplest \shelly{non-trivial} shape space; it  has    dimension \shelly{$2$}. 

\begin{figure}[t!]
\centering
\includegraphics[width=.7\textwidth]{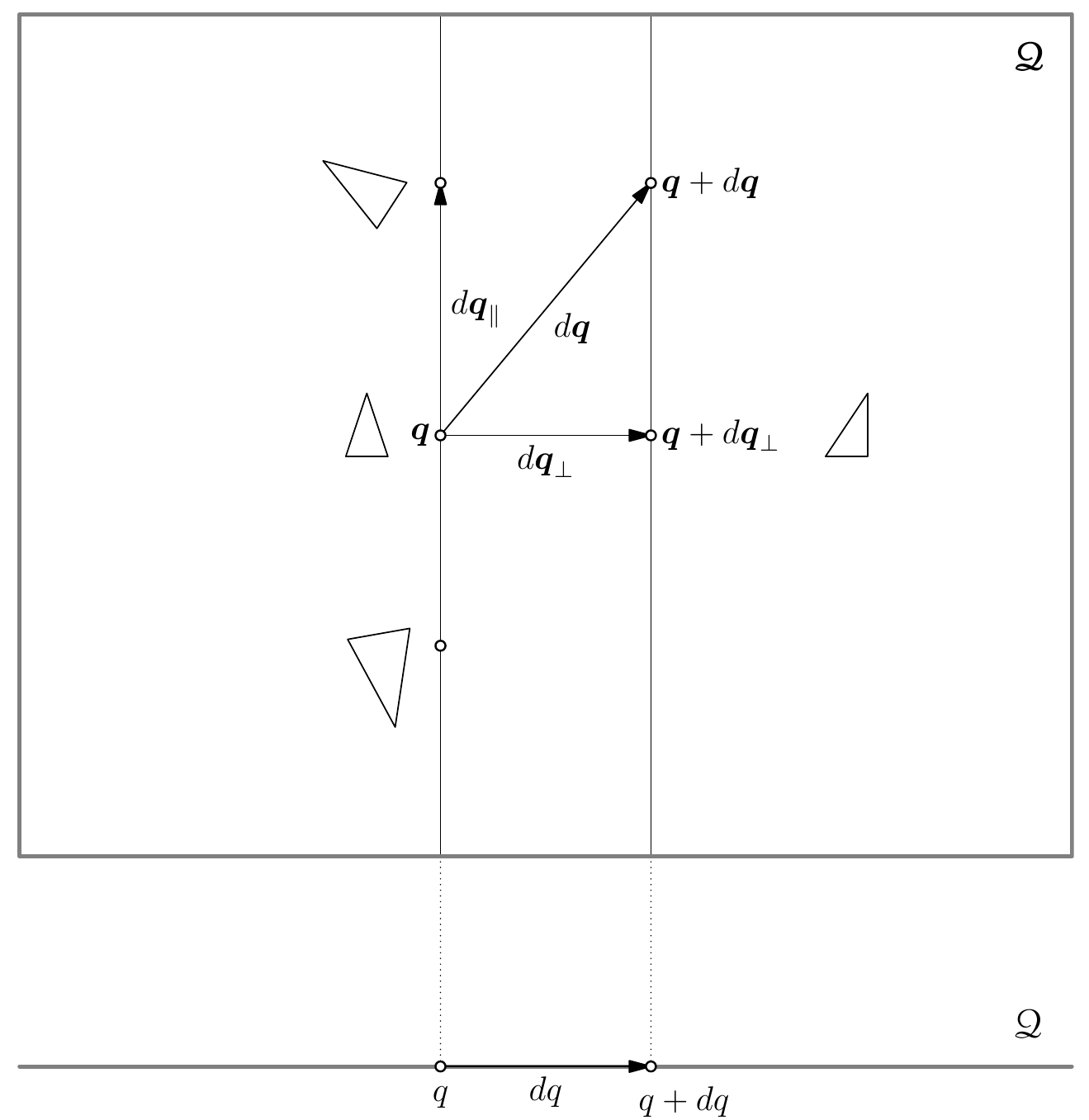}
\caption{\small Absolute configuration space $\QA$ and shape space $\QS$ (for a system of three particles). The fiber above shape $q$ consists of absolute configurations \shelly{$\q$} differing by a similarity transformation of Euclidean  space and thus representing the same shape $q$. Real change of shape occurs only by a displacement to a  neighboring fiber 
 $q+dq$. Only the orthogonal component  $d{\q }_\perp$ of $d{\q }$ represents real change, while the vertical displacement $d{\q }_{\|}$ does not contribute;  ${\q }+ d{\q }_\perp$ is the absolute configuration in the fiber above $q + dq$   closest  to $q$ in the  sense of the $g_B$-distance (best matching). }
 \label{fig1}
\end{figure}

 Accordingly, it is helpful to represent absolute configuration space $\QA$  in geometrical terms as a fiber bundle, with shape space $\QS$  as \shelly{the} base manifold and \shelly{with} the fibers generated by the the similarity group $G$ group. The points on each fiber represent physically equivalent states. See Fig. \ref{fig1}.

 \subsubsection{Metrics on Shape Space and Best Matching}
\shelly{The} topology of shape space is  well defined by its construction as \shelly{a} quotient space, but topology, of course,  does not fix  a metric. A metric should provide  more, namely a natural notion of  distance on $\CS $. And since 
each point in $\CS $ represents a
class of configurations of $N$ particles related by a similarity transformation, the distance between two elements of $\CS $ induced by the
metric
should not recognize  any absolute configurational difference due to an overall
translation, or rotation, or dilation. In other words, it 
should  provide a measure of the \emph{intrinsic} difference
between two absolute configurations (that is, not involving any consideration
regarding
how such configurations are embedded in Euclidean  space). 

It turns out \cite{QMSS}
that  {\em a metric 
on  absolute configuration space  $ \QA$ that is invariant under the group $G$ of similarity transformations  of Euclidean space, given by  a suitable ``conformal factor'' (to be explained below), defines canonically a  metric on shape space $\QS$}. 

 To understand why this is so, consider the 
representation of   absolute configuration space $ \QA$  \shelly{as} a fiber bundle  with  each fiber being  homeomorphic to $G$ and $\QS$ being its  base space (see  Fig. \ref{fig1}).   So, if $g$  is a metric invariant under any  element of $G$, the  tangent vectors at each point $\q\in\QA$    are   {\em naturally} split into ``vertical''   and ``horizontal,'' where by ``naturally'' we mean that the splitting itself is invariant under the action of $G$.  The vertical ones correspond to (infinitesimal) displacements along the fiber through $\q$ and the horizontal ones
are those that are orthogonal to the fiber, i.e., to the vertical ones, according to  the relation of orthogonality defined by $g$. More precisely, if $d\q$ is an infinitesimal displacement at $\q$, 
we have 
$$ d\q= d\q_{\|} + d\q_\perp\quad \text{with} \quad 
g (d\q_{\|}, d\q_\perp)  =0  $$ 
(see Fig. \ref{fig1}), with $ d\q_{\|} $ vertical and $d\q_\perp$ horizontal.

The corresponding Riemannian metric on $\QS$ is defined  as follows.  Let $\qs$ be a shape,  $\q$ \shelly{any} absolute configuration in the fiber above $\qs$, and $d\q$ \shelly{any }displacement at $\q$. Since $g$ is invariant under the group $G$,  the  length  of $d\q_\perp$  has the same value for all absolute configurations $\q$ above $\qs$.  \shelly{Thus} we may set the   length  of $d\qs$   equal to that of $d\q_\perp$ and hence obtain the  Riemannian  metric $g_B$  on  $\CS$   
\begin{equation}
g_B  (d\qs, d\qs)  =   g  (d\q_\perp, d\q_\perp)    \,.
\label{eq:bmetr}
\end{equation}

The subscript $B$ stands for Barbour and Bertotti, as well as  base and best matching.  In fact, The distance on
$\CS$ induced
by $g_{B}$ is exactly the one  resulting from applying  Barbour's best
matching procedure. Consider two infinitesimally close shapes, $\qs$ and $\qs +
 d\qs$, and let $\q$ be any absolute representative of $\qs$, i.e., any point in
the fiber above $\qs$. The $g_B$-distance between these shapes is then given
by the $g$-length of the vector $d\q$ such that (i) $d\q$ is orthogonal to the
fiber above $\qs$ and  (ii) $\q + d\q$ is an absolute representative of $\qs +
d\qs$. It follows that $\q + d\q$ is the absolute configuration closest to $\q$ in
the fiber above $\qs +
 d\qs$. Thus the $g_B$-distance is the ``best matching'' distance.

 \subsubsection{Invariant Metrics on Absolute Configuration Space}
\shelly{So we need} to show how to construct an invariant metric \shelly{$g$} on $\QA$.   
 Let  $g_e$ be the mass-weighted Euclidean metric on $\QA$ with positive weights $m_ \alpha $, $ \alpha =1, \ldots, N$, 
(the masses of the particles),
in particle coordinates $\q = (\vect{q}_1, \ldots, \vect{q}_\alpha, \ldots, \vect{q}_N)$  given by 
 \begin{equation}
ds^2  =  \sum_{ \alpha =1}^N    m_\alpha d\vect{q}_ \alpha  \cdot   d \vect{q}_ \alpha   \,,
\label{eq:euclid}
\end{equation}
i.e., with 
$[g_e]_{ij} = m_{\alpha_i}  \delta_{ij}$,  where the $i$-th component 
 refers to the $\alpha_i$-th particle. The corresponding line element is
  \begin{equation}
|d\q |= \sqrt{ \sum_{ \alpha =1}^N    m_\alpha d\vect{q}_ \alpha  \cdot   d \vect{q}_ \alpha }  \,.
\label{eq:euclidmetric}
\end{equation}
%
The metric defined by \eqref{eq:euclid} is invariant under rotations and translations, but not under a dilation  $\q  \to\lambda \q $, where $\lambda$ is a positive constant. 
Invariance under dilations is achieved
by multiplying $|d\q|^2$ by a   scalar function $f(\q)$ that is invariant under rotations and translations and  is homogeneous   of degree $-2$. We call $f$ \shelly{a} {\em conformal factor}. So, for any choice of  $f$,   \begin{equation} 
g = f g_e\,, \quad\text{i.e,}\quad 
g (d\q, d\q) = f(\q) |d\q|^2 \,,
\label{eq:confmetric}
\end{equation}
is an invariant metric on $\QA$, yielding
the metric on  shape space
\begin{equation}
g_B(dq, dq)  =  f(\q)  |d\q_\perp |^2  \,.
\label{eq:bmetr1}
\end{equation}
For the associated  line element  we shall write
 \begin{align} \label{eq:bmetrxxx1}
 ds = |dq| =\sqrt{g_B(dq, dq)  } =\sqrt{f(\q)} \, |d\q_\perp |  \,. 
 \end{align}

 \subsubsection{Conformal Factors}\label{conformalfactors}

Many choices of conformal factors  are possible. 
One that  was  originally suggested by
Barbour and Bertotti is\footnote{Here and in the following examples the conformal factors   are modulo   dimensional factors.}
\begin{align}f(\q) =f_a(\q)\equiv \left(\sum_{\alpha < \beta}
\frac{m_\alpha m_\beta}{|\vect{q}_{\alpha}-\vect{q}_{\beta}|}\right)^{2}\,. \end{align} 
Another example is \begin{align}  f(\q) =f_b(\q)\equiv {\rr ^{-2}} \,, \end{align} where
\begin{align}  \rr^2 &= \sum_\alpha m_\alpha{ \ovq_\alpha}^2 
=  \frac{1}{\sum_\alpha m_\alpha} \sum_{\alpha< \beta} m_\alpha m_\beta | 
\vect{q}_\alpha  - \vect{q}_\beta |^2  
 \label{mominsca}
\end{align}
with $ 
\ovq_\alpha = \vect{q}_\alpha - \vect{q}_\text{cm}\,,
$
the coordinates relative to the center of mass
\begin{align}
\vect{q}_\text{cm}  = \frac{\sum_\alpha m_\alpha\vect{q}_\alpha}{\sum_\alpha m_\alpha}\,.
 \end{align}
$I\equiv \rr^2 $ is sometimes called (but the terminology is not universal) the 
moment of inertia of the configuration $\q$  about  its center of mass. This quantity
 is half the trace of the moment of inertia tensor $\mathsf{M}$,
 \begin{align}   \rr^2= \frac12  \Tr  \mathsf{M} \,. \end{align}
 We recall that 
$ 
 \mathsf{M} = \mathsf{M} (\q) 
$, the  tensor  of inertia of the configuration $\q$ about any orthogonal cartesian system $x$,$y$,$z$  with origin  at the center of mass of the configuration $\q$,  has  matrix elements  given by the standard formula
\begin{align}\label{mateltensin}
M_{ij}= \sum_{\alpha=1}^N m_\alpha (\rrr_\alpha^2\delta_{ij} - \rrr_{\alpha i}  \rrr_{\alpha j}  )\,,
\end{align}
where $i,j=x,y,z$,  $\rrr_{\alpha x} \equiv  x_{\alpha} $,  $\rrr_{\alpha y} \equiv y_{\alpha} $, 
$\rrr_{\alpha z}  \equiv  z_{\alpha} $, and $\rrr_\alpha^2=  x_{\alpha}^2 +  y_{\alpha}^2 +z_\alpha^2$.

A  choice of conformal factor that has not been considered in the literature  is 
\begin{align}\label{pconfacty}
f(\q) =f_c(\q)\equiv   \rr^{-\frac{8}{7}} (\det\mathsf{M})^{-\frac{1}{7}}\,.
\end{align}
Since $\det\mathsf{M}$ scales as $\rr^6$, $f(\q)$ given  by \eqref{pconfacty}  scales as it should, namely, as  $\rr^{-2}$. 
Though at first glance this choice \shelly{seems  unnatural, it is in fact very} natural once the motion of shapes  is analyzed from a quantum perspective \cite{QMSS}.

Finally, we  give other   two examples:
\begin{align}\label{cfd}
f(\q) &=f_d(\q) \equiv  \sum_{\alpha < \beta}
\frac{m_\alpha m_\beta}{|\vect{q}_{\alpha}-\vect{q}_{\beta}|^2} \\
\label{cfg}
f(\q) &=f_g(\q) \equiv \rr^{-1} \sum_{\alpha < \beta}  \frac{m_\alpha m_\beta}{|\vect{q}_{\alpha}-\vect{q}_{\beta}|} \,.
\end{align}
The first one corresponds to a natural modification of the Newtonian gravitational potential
and the second,  discussed in \cite{barbour2003scale}, corresponds to a dynamics very close to that of Newtonian gravity.

\subsection{Free Motion on Shape Space}
All one needs to define the simplest  motion on shape  space, be it classical or quantum (Bohmian), is the metric $g_B$ on shape space. In the classical case,
such a metric $g_{B}$ directly yields a law of {\em free motion} on shape space, 
that is,
geodesic motion with constant speed.

In the quantum (Bohmian) case, the situation is similar. Equations  \eqref{eq:bmonmany1x} and
 \eqref{eq:bmonmany2x} define  immediately free quantum motion on shape space with Riemannian metric $g= g_B$ as 
the motion on shape space 
given by  the evolution equations 
\begin{align}\label{eq:bmonmany1}
\frac{dQ}{dt} &= \hbar  \, \mathrm{ Im } \frac{\nabla_{B}  \psicap{}}{\psicap{}} \\
\label{eq:bmonmany2}
i\hbar \frac{\partial \psicap{}}{\partial t} &= -\frac{\hbar^2}{2} \Delta_B \psicap{} \end{align}
where  $\Delta_{B}$ and  $\nabla_{B} $ are,
respectively, the Laplace-Beltrami operator and the gradient on the
configuration space equipped with the Riemannian  metric \eqref{eq:bmetr1}.

We have insisted on free motion, because it is the simplest motion on shape space,
which has, surprisingly,  a very rich structure. Of course, one may introduce a potential term both in the classical and the quantum case in a straightforward way: (minus) the gradient of the potential in the right hand side of the geodesic equation for the classical case and the usual potential term in the right hand side of Schr\"odinger's equation.

This is all there is to say about the {\em formulations} of classical and quantum theories on shape space, modulo a caveat: so defined, classical and quantum motion rely on an external absolute  time, which, from a relational perspective,  is \shelly{problematical},  see below.

\subsection{Fundamental Level vs. ``Human'' Level}

Let us draw some morals from the foregoing. First of all, it is rather clear that   to define motion on shape space is very \shelly{easy}. The hard work,    basically in the quantum case \cite{QMSS},  involves the transition from the fundamental level of shape space to what could be called  the ``human'' level of absolute configuration space. We say ``human'' because it is the description we usually use. Even if shape space is fundamental, we humans don't typically  formulate things in terms of shape space. \shelly{That  doesn't come quite so naturally} to us, even if shape space is in a deeper sense   more fundamental  and natural. Traditionally we do things in this way for  good human reasons and probably for many others.  

The level of absolute configuration space is a sort of emergent level because \shelly{it} is not fundamental. How things are described at that level is imposed by us.  It is ``human'' in the sense that it  is based on our choice of description. This is why we consider it a Kantian component of physical theory.  As we shall elaborate below,  much of what \shelly{we've taken} as fundamental is not fundamental, but imposed by us by \shelly{a} proper choice of gauge.  \shelly{Interactions provide an important example.}

\subsection{Emergence of Interactions}
 Given  \shelly{a} motion in shape space, there is a huge host of motions in absolute space that are compatible with it, the only constraint  being  that they should project down to \shelly{the} motion  in shape space.  In other words, recalling the fiber bundle structure of absolute configuration space described above, we have that
a representation in absolute configuration space of the motion in shape space is  given by  a ``lift'' of the motion from the base into  the fibers. 

Such lifts can rightly be called gauges. In the classical case it turns out that in some gauges the law looks unfamiliar but there is (at least) one gauge in which,  after performing a time change---representing indeed another gauge freedom when also time is seen as relational---the law of motion is Newtonian with a potential appearing. 
We call it  the {\em Newton gauge}. 
The potential depends on the choice of the invariant metric \eqref{eq:confmetric} in absolute configuration space,  with the  various possibilities for the conformal factor, some of them listed  at the end of Section \ref{sec: spafcs}. 

More or less the same is true for the quantum case,  where however the gauge yielding  ordinary Bohmian mechanics in absolute configuration space---which we call the {\em Schr\"odinger gauge}---emerges only for a stationary, i.e. time-independent, wave function (such as with the  Wheeler-DeWitt  equation) on shape space. This again \shelly{involves} regarding time as being relational, with an external absolute time playing no physical role.   

Also here, while the fundamental  physics is given by a free Bohmian dynamics in shape space, in the Schr\"odinger gauge potential terms appear. One potential term is determined by the scalar curvature induced by the  invariant metric on absolute configuration space. Another potential term arises from the gauge freedom we have to lift the Laplace-Beltrami operator from shape space to absolute configuration space, where \shelly{more} gauge freedom arises from allowing transformations  of the lifted wave function.

\section{Sources of Gauge Freedom}
The main sources of gauge freedom are a {\em lift}   from the base into  the fibers \shelly{and}
a {\em time change}, both for the classical and quantum case. For the quantum case 
there is in addition a gauge transformation
\begin{equation}  \Psi \mapsto \Psi' = F\Psi\,, \quad\text{ with $F$ real.}
\label{eq:qgt}
\end{equation}
The implications of this will be discussed below in detail.
\subsection{Classical Gauges}
In the classical case, consider the {free motion}  $\Qs=\Qs(t)$ on shape space, that is, the geodesic motion with constant speed.
A very natural   lift   of  this  motion in  absolute configuration space is its  {\em horizontal lift}, that is,  a motion $\Q= \Q(t)$ in absolute configuration space  that  starts at  
some  point $\q_1$ on the fiber above $\qs_1$  and is {\em horizontal}, i.e., the infinitesimal  displacements  $d\Q$ are all horizontal. (Note that  the final point $\q_2$ in the fiber above  $\qs_2$ is then uniquely determined.)  
 We call this    the {\em invariant \shelly{(or horizontal)} gauge}. 
 
 It can be shown \cite{QMSS} that in this gauge, the familiar 
total momentum $\vect{P}$,  total angular momentum $ \vect {J}$ and (maybe less familiar)  dilational momentum $ D$ are all zero,  i.e.,
\begin{eqnarray}
 \vect{P}&=&\sum_{ \alpha =1}^N m_ \alpha \frac{d \vect{Q}_ \alpha }{d t} = 0\label{constraint1}\\
 \vect{J}&=&\sum_{ \alpha =1}^N  m_ \alpha  \vect{Q}_ \alpha \times \frac{d \vect{Q}_ \alpha }{d t} = 0\label{constraint2}\\
{D}&=&\sum_{ \alpha =1}^N  m_ \alpha  \vect{Q}_ \alpha \cdot \frac{d \vect{Q}_ \alpha }{d t} = 0\label{constraint3}\,.
\end{eqnarray}

 \subsubsection{The Newton  Gauge}
By a suitable change of speed, that is by a suitable time change $t' = t'(t)$, namely,
\begin{align}\label{timechange}
\frac{v}{\sqrt{f}  }  \frac{dt}{dt'} =  \sqrt{2f} \,, \quad\text{i.e.,} \quad   \frac{dt'}{dt} = \frac{v}{\sqrt{2} f} \,,
 \end{align} we  \shelly{obtain} another gauge that  we call the {\em Newton  gauge}, a gauge in which  the motion is  Newtonian, i.e., it satisfies Newton's equation $F=ma$ for suitable $F$. More precisely, \shelly{the particle} positions $\vect{Q}_ \alpha$, $\alpha=1, \ldots, N$,  forming the configuration $\Q$ \shelly{then} satisfy Newton's equations
\begin{align} \label{neteqab}m_\alpha \frac{d^2\vect{Q}_ \alpha }{d {t' }^2} = -\vect{\nabla}_\alpha V (\vect{Q}_ 1, \ldots \vect{Q}_ N )\,. \end{align}
with $V(\q)= -  f (\q) $, the conformal factor.

In this regard, one should observe that if time is relational,  changes of speed, such as that given by  \eqref{timechange}, provide equivalent representations of the same motion. Accordingly, the use of one time variable instead of another is a matter of convenience, analogous to the choice of a gauge. 
The  choice of time variable that leads from the invariant gauge to  
 the Newton gauge  is  
the gauge fixing condition
\begin{align}\label{myspeed} E= \frac12 \left|\frac{d\Q}{dt'}\right| ^2 +V =0 \,, \end{align}
 that is, that the total Newtonian energy be zero. Of course, other gauges could be useful, as discussed next.
 
 \subsubsection{The Expansion Gauge}
 Another gauge that could be useful is what could be called the {\em expansion gauge}, a gauge in which you do not \shelly{use a horizontal lift, but instead have a uniform expansion of configurations  taking place}---in the horizontal gauge the total overall size does not \shelly{change: the universe would not expand since  the total dilational momentum is zero in that gauge\cite{barbour2003scale}. In contrast, in the expansion gauge   configurations} and thus the universe would expand.  Why one would consider such a gauge? One reason is that in this way one  might simplify the form of the gravitational forces (see below), so that  they would become more familiar in this gauge. In this regard, note that
 since the motion is not horizontal, after suitable time change the Newtonian form of the equation of motion would involve  other terms, in addition to the forces generated by the conformal factor. Another reason is that one would obtain in this way  a striking explanation of why we think that the universe is expanding.

 \subsubsection{A Note on Newtonian Gravitation}
 \label{newgra}
We have seen  that in the Newton gauge, when the physical law on shape space is free motion (or even non-free motion), the potential $V=-f$ appears, where $f$ is the conformal factor. We mentioned some choices for $f$ in Section \ref{conformalfactors}. No such choices,  which are necessarily functions homogenous of degree $-2$, seem to
yield exactly the Newtonian gravitational potential $U_g$.  While we believe  the  detailed exploration of the implications of the models discussed here is worthwhile, we nonetheless  regard the models explored in this paper, both classical and quantum, as  toy models,  so that  such an analysis of them, with the expectation of recovering well established physics, might be somewhat inappropriate or premature.  

However,  it should be observed that some of the conformal factors given in Sect. \ref{conformalfactors}, e.g., $f_a$ and $f_g$,   indeed give rise to  a force law  in the Newton gauge that is very close to that of  the Newtonian gravitational \shelly{force}.  Note for example  that  for  the conformal factor $f_g$ the corresponding potential   is of the form $V_g = \rr ^{-1} U_g$,   where $U_g$ is the  Newton gravitational potential and   $\rr $ in the Newton gauge is a constant of the motion. The 
force arising from this  potential adds to the  Newtonian force a very small  centripetal  correction that allows $I = \rr^2$,  the moment of  inertia  about  the center of mass,  to remain constant   \cite{barbour2003scale}. It is worth exploring these possibilities in the expansion gauge.

\subsection{Quantum Gauges}\label{scg}
Let $\Qs=\Qs(t)$ be a Bohmian motion in shape space, that is,  a solution of \eqref{eq:bmonmany1} with the wave function  $\psicap{} $ being a solution of Schr\"odinger's  equation  \eqref{eq:bmonmany2} on shape space. \shelly{As} in the classical case, we wish  to characterize motions in absolute space that are compatible with motions
in shape space, that is,   motions  
$\Q =\Q(t)$ in $\QA$  that project down to $\Qs=\Qs(t)$ in $\QS$, i.e., such that
\begin{align}  \pi (\Q(t)) = \Qs(t) \,, \end{align}
where $\pi$ is the canonical projection from $\QA$ \shelly{to} $\QS$.
Clearly, there  are  a great many possibilities  for compatible motions in absolute configuration space.

As in the classical case, one may restrict the possibilities by considering natural gauges. And as in the classical case, where one  looks for gauges such that the absolute motions satisfy Newton's equations,
in the quantum case we now look for gauges such that the  compatible motions on  $\QA$ are themselves Bohmian motions, i.e. motions generated by  a  wave function in the usual sort of way.

\subsubsection{Three Quantum Gauges}
Suppose that  we proceed as in the classical case and   take   a  horizontal lift    of a  motion $\Qs=\Qs(t)$ in shape 
space, that is,  an absolute  motion for which  the infinitesimal  displacements  $d\Q$ are all horizontal. Let us now consider the lift  to  $\QA$  of  a  wave function  $\psicap{}$  on $\QS$, namely, the wave function $ \firstgaupsi$ on absolute configuration space such that
\begin{equation}\label{strhl}  \firstgaupsi(\q) = \psicap{}(q)  \end{equation} for any point $\q$ on the fiber above $q$.
Let   $\nabla_g$ be the gradient with respect to the invariant measure  \eqref{eq:confmetric}. Then
the vector
$\nabla_g  \psibh(\q)$ in $\QA$  is  horizontal and  the motions on $\QA$  defined by
\begin{align}\label{velogB}  \frac{d\Q}{dt}= \hbar  \, \mathrm{ Im } \frac{\nabla_{g} \psibh}{\psibh} \end{align}
are  horizontal lifts of motions on  $\QS$.
So, in the quantum case, horizontality is immediate.

Let us now consider the time evolution of the   lifted wave function $\psibh $ on  $\QA$.  Let  $\lDeltag$  be  a lift to absolute configuration space  of the Laplace-Beltrami operator $\Delta_B$ on shape space, namely an operator on $\QA$ such that
\begin{align}    \lDeltag \psibh  =  \Delta_B \psicap{}  \,. \end{align}
Then
\begin{align}\label{hamsfac}  i\hbar \frac{\partial\psibh}{\partial t} = \firstgauH \psibh\,,  \quad
\text{with}\quad
\firstgauH = -\frac{\hbar^2}{2}  \lDeltag  \,.\end{align}
It might seem   natural to guess that  $\lDeltag$   coincides  with $\Delta_{g}$,
the  Laplace-Beltrami operator  with respect to  $g$,   but this  is wrong;
nor is  $\bHabh_1$   a  familiar sort of Schr\"odinger Hamiltonian, with or without a potential term.

While  $\psibh $ need not obey any familiar Schr\"odinger-type equation, one may ask whether  there exists a gauge equivalent wave function that does. By gauge equivalent wave function we mean this:  If one writes  $\psibh $ as $Re^{(i/\hbar) S} $ one sees that the velocity field given by \eqref{velogB}  is just $\nabla_g S$, so transformations of the wave function
\begin{align}  \label{psiqtransf}  \psibh\to\psibh '  =  F\psibh\,, \end{align}
where $F$ is a positive function, do not change its phase and thus the velocity. 

It turns out that there exists a positive function  $F$ such that  $ \thirdgaupsi= F \firstgaupsi$   satisfies  a Schr\"odinger-type  equation on absolute  configuration space for a suitable potential $V$, namely,  
\begin{align}\label{hamsf2}  i\hbar \frac{\partial \psibb}{\partial t} = \bHabbb \psibb   \end{align}
with
\begin{align} 
\label{eq:Steptwo} \bHabbb  
&= 
- \frac{\hbar^2}{2} \sum_{\alpha=1}^N  \vect{\nabla}_{\alpha}\cdot\frac{1}{ f m_\alpha }\vect{\nabla}_{\alpha} +V
\\
\label{eq:Steptwo}   &= -\frac{\hbar^2}{2} \nablab\bigcdot \frac{1}{f}\nablab +V\,,
\end{align}
%
%
where 
 $\nablab$ and 
 $\nablab\bigcdot$ are the  gradient and divergence with respect to the  mass-weighted Eucidean metric 
\eqref{eq:euclid}, i.e.,  
 \begin{align}\label{nsblab}  \nablab = \left(\frac{1}{m_1} \vect{\nabla}_{1} , \ldots,    \frac{1}{m_N} \vect{\nabla}_{N}    \right) \end{align}  
 and 
 \begin{align}\label{nsblab2}  \nablab \bigcdot= \left(\vect{\nabla}_{1} , \ldots,     \vect{\nabla}_{N}    \right) \bigcdot \, .\end{align} 
Here $f$ is the conformal factor, and
\begin{align}V &=V_1+V_2  \\
\label{V_1}
 V_1 &= 
-\frac{\hbar^2}{2}\frac{  \lDeltag J^{1/2}}{ J^{1/2}} \\
\label{exvtwo}
 V_2 &=   -\frac{\hbar^2}{2} f^{\frac{n}{4}}\Delta_g\big(f^{-\frac{n}{4}}\big)\,,
\end{align}
with
\begin{align}\label{forforJ}
J= \rr f^{7/2}  \sqrt{\det \mathsf{M}}\,,
\end{align}
where 
 $\rr=\rr(\q)$  is 
 given by  equation \eqref {mominsca} 
 and $\mathsf{M} = \mathsf{M} (\q) $  is the tensor  of inertia of the configuration $\q$ about any orthogonal cartesian system $x$,$y$,$z$  with origin in its  center of mass  and  with  matrix elements  given by   \eqref{mateltensin},  and where $ \lDeltag$ is the canonical lift  of $\Delta_B$ (see \cite{QMSS}  for details on this and a proof of the aforementioned statements).
 
 \shelly{There is a 2-gauge,  intermediate between the 1-gauge and the 3-gauge, that we shall not describe further here.}
 
 \subsubsection{The Schr\"odinger gauge}
We shall now describe what we think is appropriate to be called the  {\em Schr\"odinger gauge}, the true quantum analogue of the Newton gauge.  If we take into account that time is relational, as we should, 
the fundamental equation for the wave function on shape space is  presumably the stationary equation
\begin{align}\label{eq:bmonmany3} 
-\frac{\hbar^2}{2} \Delta_B \psicap{} = \mathscr{E} \psicap{} \,,
\end{align}
where \shelly{$\mathscr{E}$ is any  given}  fixed constant (for example $\mathscr{E}=0$). 

As before, let $\psibh$ be the lift of $\psicap{}$ to $\QA$, so that   $\psibh$ satisfies the equation
\begin{align}\label{hamsfacst}   (\bHabh_1 - \mathscr{E}) \psibh =0 \,, \end{align}
with $\bHabh_1$  a lift of $\Hab$ as in \eqref{hamsfac},
and  the evolution on the absolute configuration \shelly{space} $\QA$  is  still given by \eqref{velogB}. 
But now, for relational time,  
{motions following the same path  with different speeds
are the same motion}. So, in the formula for the 
 gradient 
on the right hand side  of \eqref{velogB}, $\nabla_g = f^{-1}\nablab$, 
we may regard  $f$ as a change of speed  defining  a new time variable that for the  sake of simplicity we shall still call $t$ (\shelly{a ``time change''}). Then
in absolute space  the guiding equation \eqref{velogB} becomes
\begin{align}\label{bohmabsolu}
\frac{d\vect{Q}_\alpha}{dt} =\frac{\hbar}{m_\alpha} \, \mathrm{ Im } \frac{\vect{\nabla}_{\alpha} \psibh}{\psibh} \,.
\end{align}



Again, $\psibh $ need not obey any familiar stationary Schr\"odinger-type equation. However, as before, we may exploit gauge freedom to transform \eqref{hamsfacst}  into a stationary Schr\"odinger-type equation. Indeed, we \shelly{now} have an even greater gauge freedom in changing the wave function and the Hamiltonian,  \shelly{see \cite{QMSS}}.
 In particular,  
there is a gauge, the Schr\"odinger gauge,  in which \eqref{eq:bmonmany3} becomes
\begin{align}\label{step31}
\bHabb \Phi= 0 
\end{align}
with 
\begin{align}\label{step312}
\bHabb= -\frac{\hbar^2}{2} \nablab^2 + U\,,
\end{align}
where $\nablab^2= \nablab\bigcdot \nablab$ is the  mass-weighed Euclidean Laplacian, and
\begin{align} \label{step32}
U= f (V_1  -\mathscr{E})   - \frac{\hbar^2}{8} \frac{n-2}{n-1} f  R_g  \,,
\end{align}
where $R_g$ is the scalar curvature of the invariant  metric~$g$ and $n=3N$.

\section{Lessons from Relational Physics}
In this closing section we shall \shelly{further consider}  what we can learn from relational physics. 

\subsection{Local Beables \shelly{and} Primitive Ontology}
We have stressed elsewhere the importance of \shelly{certain} local beables, what we  call the primitive ontology of a theory---stuff in space evolving in time---which is what the theory is fundamentally about. Relational physics \shelly{requires} a more flexible understanding of this notion.

\shelly{To appreciate} this point, note that for relational space the state of the universe at a particular location is not, in and of itself, meaningful. In that sense, for \shelly{relational} physics, there are no local beables \shelly{in the usual sense, and} locality itself can't  easily be meaningfully formulated. 

\shelly{Similarly} one can't meaningfully consider the behavior of individual particles without reference to other particles, since there is no absolute space in which an individual particle could be regarded as moving. And even for a pair of particles, to speak meaningfully of the distance between them, a third particle would be required, to establish a scale of distance. And similarly for galaxies.

\shelly{Relational physics is, in fact,} genuinely holistic, and suggests the holistic character of quantum physics associated with entanglement and quantum nonlocality.

\subsection{Projectivity}

\shelly{Compare the guiding equation \eqref{Bohm} with the same equation but with the denominator $\Psi^*\Psi$  omitted.
 The latter simpler equation might be regarded as in some sense more fundamental, with \eqref{Bohm} itself regarded as arising from a convenient choice of gauge corresponding to the freedom of time-change associated with relational time.}
 
The time-parameter corresponding to the use of the denominator in \eqref{Bohm}  has the nice feature that the dynamics using that time-parameter depends on fewer details of the wave function than would be the case if the denominator were deleted: with the denominator the dynamics depends only on the ray of $\Psi$, with $\Psi$ and $c\Psi$ yielding the same dynamics for any constant $c\neq0$. This has a particularly nice implication for the behavior of subsystems. 

\shelly{With} this choice of time-parameter the dynamics for a subsystem will often not depend upon the configuration of its environment, with the subsystem evolving according to an autonomous  evolution involving only  the configuration and the (conditional) wave function of the subsystem itself \cite{durr1992quantum}.
This would happen when the subsystem is suitably decoupled from its environment, for example for a product wave function when there is no interaction between system and environment. Without the denominator \shelly{this} would not be true, and there would appear to be an additional  nonlocal dependence of the behavior of a subsystem on that of its environment that would not be present with a time-parameter associated with the use of the usual denominator.

\subsection{Non-Normalizability of the Wave function}\label{nnwf}
Note that since it is translation and scaling invariant, the wave function in the \Sc\  gauge or in any of the  gauges discussed in Sect.\ref{scg}) are not normalizable.
In other words, the corresponding $|\widehat{\Psi}|^2$ measures are   non-normalizable. 
However, since the non-normalizability arises from 
unobservable (and, from a shape space point of view, unphysical) 
differences and dimensions it should somehow not be a problem. 
%

Nonetheless, the real question is how  the empirical 
distributions arising from the fundamental shape space level \shelly{are related to}  those 
coming from the physics in a gauge. While the different gauges, such as the \Sc\  
gauge, correspond to theories that, we argued,  are empirically equivalent to 
the fundamental shape space theory,  that was only in purely dynamical terms. 
We have not yet addressed the possible differences in empirical distributions 
that may arise. We would like to see that they don't, \shelly{i.e., that probabilities naturally associated with relational physics on shape space yield familiar Born-rule probabilities when the physics is lifted to absolute configuration space.}

There are several considerations that suggest that the non-normalizability should not be a genuine problem. First of all, 
 as just mentioned, the non-normalizability arises only from non-observable dimensions, suggesting that it should be physically  irrelevant.
Moreover, it is the universal wave function $\gaupsi$ (in any of the   gauges) that is not normalizable. But the  universal wave function is rarely used in practice. In quantum  mechanics we usually deal, not with the entire universe, but with small subsystems of the universe. The wave functions with which we usually deal are thus conditional wave functions, and there seems to be no reason why these should fail to be normalizable.

As already said,  non-normalizable wave functions  tend to occur in quantum cosmology. Such wave functions would normally be regarded as problematical and unphysical (since the formal structures of orthodox quantum mechanics, with their associated probabilities, are crucially based on the notion of a Hilbert space of square-integrable, i.e. normalizable, wave functions).
However, for our analysis  in \cite{QMSS}  the   nonnormalizability of  the measure on configuration space turned out to be crucial \shelly{for its success} (see below). Hence what  from an orthodox perspective is a vice is transformed into a virtue in relational Bohmian mechanics.
\subsection{Conditional Wave Function and Path Space}
To substantiate the last statement we mention a few ingredients  of our analysis in 
\cite{QMSS}, without any pretension of completeness.
 \subsubsection{The Conditional Wave Function}
The very notion of conditional wave function \eqref{conprobfor} presupposes 
the possibility of \shelly{expressing the configuration $Q$ of the universe as a pair $(X,Y)$ consisting of the configuration $X$ of a system, the $x$-system, and the configuration  $Y$ of its environment (the $y$-system).} The holistic character of shape space physics emphazized above
does not allow for \shelly{this}, that is,  there is no natural product structure
$$\QS=\QS_{sys} \times \QS_{env}$$ for shape space: Here  the system is a collection of (labelled) particles with its own shape space $\QS_{sys}=\mathcal{X}$, the set of possible shapes $X$ of the system, and the environment consists of the rest of the particles of the universe, with shape space $\QS_{env}=\mathcal{Y}=\{Q_{env}=Y\}$, with $Y$ the shape associated with the particles (labels) of the environment.
The crucial fact is that it is not true that 
$$\QS=\mathcal{X} \times \mathcal{Y}.$$
$X$ and $Y$ don't involve sufficient information to determine the complete shape $Q$. What is missing is the spatial relationship between these shapes. 

Nonetheless we have that $\QS$ can be identified with
$\mathcal{X}_Y\times_y\mathcal{Y}=\{(X,Y)|Y\in\QS_{env}, X\in\QS_Y\}\,,$
where
$\QS_Y=\{Q\in\QS\,|\, Q_{env}=Y\}.$
We may then define the conditional wave function for the subsystem, for $Y\in\QS_{env}$ and universal wave function  $\Psi$, by
\begin{equation}\label{cwfu}\psi(x)=\Psi(x,Y), \quad x\in \QS_Y. \end{equation}
This looks like the usual conditional wave function, but it is important to bear in mind that, unlike with the usual conditional wave function, here $x$ represents the shape of the universe for a fixed $Y$ and there are no obvious natural coordinates to efficiently describe it. 

The conditional wave function \eqref{cwfu} can be lifted to absolute configuration space and one has the formula
\begin{equation}\label{psihat2} \liftw{\psi}(\lift{x})=\liftw{\Psi}(\lift{x}, \lift{Y}), \end{equation}
Here the lifts 
$\lift{x}$  of $x$ and 
 $\lift{Y}$  of $Y$ involve something akin to the choice of a frame of reference in absolute space, and the ``hat'' on $\Psi$ refers to any of the gauge  equivalent representations   of the universal wave function   that  we have  described in Section  \ref{scg}.

 \subsubsection{The Conditional Probability Formula}

One might wonder whether something analogous to \eqref{condprob} holds true.
The answer is yes, but with several provisos.

It turn out that in the gauges 1 \shelly{and} 3 described in Section  \ref{scg}, indeed something similar holds. More precisely, in \shelly{the} 3-gauge the conditional distribution is exactly given by the modulus square of the conditional wave function  \eqref{psihat2} as in \eqref{condprob}. The proof of this is  rather intricate and lengthy\shelly{\cite{QMSS}}.

In the  \Sc\  gauge, however, a conceptual issue arises as to exactly what of physical significance this conditional distribution represents.  After all,   the transition to the \Sc gauge required relational time, but if we take relational time seriously,
what is physical is not the configuration $Q_t$  of the universe at some time $t$, but the geometrical path of the full history of the configuration, with no special association of the configurations along a path with times.
In this (more physical) framework, the conditional distribution of the configuration $X_t$ of a subsystem given the configuration $Y_t$ of its environment is not meaningful.

What is meaningful is (i) a probability distribution  $\Ppath^\Psi$ on the space $\mathscr{P}$ of (geometrical, i.e. unparametrized) paths  and (ii) the conditional distribution relative to $\Ppath$ of the configuration $X_Y$ of the subsystem when the path $\gamma\in \mathscr{P} $ has environmental configuration $Y,$ {\it given} that the path passes through a configuration with environment $Y$, $Y \in \gamma$.
In this more general framework, one arrives \shelly{at} a conditional probability formula analogous to to \eqref{condprob}, namely,
\begin{equation}\label{cp1}
\Ppath^\Psi(X_Y\in dx|Y\in\gamma)  =    | \liftw{\psi}_{S}(\lift{x})|^2 d \lift{x} \,,
\end{equation}
where the subscript $S$ refers to the \Sc\  gauge. Also the proof of this theorem is  rather intricate and lengthy. 
An assumption of this theorem is what we have \shelly{called} in \cite{QMSS} the ``existence of   clock variables'' 
for the Bohmian  dynamics, which a mild assumption for a sufficiently structured universe.

 \subsubsection{Non-Normalizable Measures} 

A crucial ingredient of our analysis is  the connection between a measure on path space and a  (stationary) non-normalizable measure on configuration space. By the very nature of this connection   the measure on configuration space and its associated wave function in fact  has to be  non-normalizable---just to
 rehearse
the point made at the end of Section \ref{nnwf}  that the nonnormalizability of the universal wave function is good!

 \subsubsection{The Physics of Sub-Systems}

On the absolute configuration space level the dynamics and the probabilities for subsystems should be of the usual form. While it is true that on the universal level the connection between $|\Psi|^2$ and probability, or, more precisely, typicality, would be broken, this would not be visible in any of the familiar every-day applications of quantum mechanics, which are concerned only with subsystems and not with the entire universe. This is the key  to approach the problem of entropy increase raised  in Section \ref{sec:wfe}.

 \subsubsection{$|\psi|^2$ vs. $|\Psi|^2$} 
 
The patterns described by the quantum equilibrium hypothesis will be typical with respect to a measure,  not on absolute configuration space, but on shape space, on the fundamental level, which is fine. There is a widespread misconception with respect to Bohmian mechanics that $|\Psi|^2$ for the universe and $|\psi|^2$ for subsystems play, physically and conceptually, similar roles. They do not, since the role of $|\Psi|^2$ is typicality while that of  $|\psi|^2$ is probability. If this distinction is too subtle, the fact that, from a relational perspective, these objects live on entirely different levels of description, $|\Psi|^2$ on the fundamental level, i.e., on shape space,  and $|\psi|^2$ on absolute configuration space, might make it easier to appreciate how very different they are.

\subsection{More Relational Points}

\subsubsection{\shelly{Identical Particles}}
There is one rather conspicuous relational aspect that we've ignored. For indistinguishable particles we should have taken one further quotient and enlarged the similarity group $G$ to include the relevant permutations of particle labels. Presumably, this would not have any relevant implications in the classical case, but  in the quantum case would lead to the correct quantum  description  of the dynamics of bosonic and fermionic shapes \cite{mellini}.

\subsubsection{Different \shelly{Ontologies}}
We do not regard particles as the \shelly{inevitable} ultimate building blocks of the universe. Far from it. Our insistence on particles is mostly methodological, not metaphysical. The point is that the particle ontology
 already suffices
to highlight the relevant  conceptual issues of relational physics. 
Our analysis---in particular that concerning the wave function of the universe, subsystems, and probability---really depends only on rather general qualitative
features of the structure of quantum relational physics, not on the details of any specific ontology.

\subsubsection{Geometry}\label{sec:geo}

We have assumed  that the geometrical relations among the  particles \shelly{are}  those of Euclidean geometry. \shelly{But we know  that the geometry of space need not be Euclidean and may not be fixed at all, but be} dynamically determined by the matter content of the universe.

\shelly{One is thus naturally lead to geometrodynamics, 
the dynamics of 3-geometries. According to
John Archibald Wheeler,  superspace, the  totality $\QS$ of 3-geometries, should be regarded as the true arena of general relativity} (see, e.g., \cite{MTW}). 
 
In many respects, superspace is analogous to shape space. The totality  $\QA$ of Riemannian metrics on 3-space is analogous to absolute configuration space, and the group $G$ of diffeomorphisms of 3-space is analogous to the \shelly{similarity group}  of Euclidean space. The fiber bundle    structure 
$ \QS= \QA/G$ (modulo singularities) has been highlighted by Bryce DeWitt \cite{dewitt1970spacetime}.

\shelly{Now DeWitt in \cite{dewitt1970spacetime} and many others after him have regarded 
 four-dimensional spacetime  as fundamental and  the dynamics of 3-geometries as gauge,  arising from  the gauge freedom  given by the so-called lapse and shift functions.
 However, one could  take the opposite view and 
consider   the 3-geometry  as fundamental  (analogously to a shape of particles) and four-dimensional spacetime as arising in a suitable gauge from a timeless history of the universe, i.e., a path,}  in superspace. This view, that has been explored by Barbour and collaborators \cite{barbour2002relativity}, leads to the conclusion
that relativity (general and special)  need not be regarded  as fundamental, but can \shelly{be regarded as arising}  from a suitable choice of gauge.
Again we encounter here \shelly{a}  striking manifestation of \shelly{a} Kantian component of physical laws.

\subsubsection{Larger Groups}
For the case of an $N$-particle universe, we have already indicated that a natural extension is to   include in $G$ the relevant permutations of particle labels.  Are there other natural extensions? How \shelly{far can the gauge group $G$}  be extended? 

In the case of 3-geometries, Barbour and collaborators have explored the possibility of extending the group of diffeomorphism \shelly{to}   include the  conformal transformations of the 3-geometries. One  arrives in this way at a conformal superspace \cite{barbour2012shape,gomes2011einstein, mercati2018shape}. They perform some massaging to arrive at general relativity (in a suitable gauge).
But maybe there is no need of any massaging, nor any need of arriving exactly at general relativity---to recover it as an approximation of a more fundamental theory is \shelly{sufficient}.

\subsection{The Quantum Gauge?}
Is there a gauge that could be called the {\em quantum gauge}? We do not know the answer, but we find that the question itself is very interesting.  It  is related to the ``why''-question we asked at the end of Section \ref{sec:wvsw}.
 
 Let us explain what we mean.
We started with the Bohmian motion on shape space governed by 
the universal wave function $\Psi$. Everything we \shelly{discussed}   in the quantum case had as a basic ingredient the universal wave function $\Psi$.     But one \shelly{should} contemplate the possibility of  starting with something that is not really quantum, not a motion guided by \shelly{a suitable} $\Psi$. 

Suppose we have a motion in shape space  that is \shelly{not of the usual Bohmian form, with no wave function $\Psi$}  involved in defining the  law of  motion of the configuration. In other words,  \shelly{suppose we discover a natural motion on shape space that  is not quantum at all.  But suppose also} that  when one takes into account the  freedom of time change \shelly{and} the freedom of how the motion is lifted to absolute space, that is,  that when one exploits all the gauge freedom we have discussed above, it turns out that the law of motion can indeed be \shelly{cast into} the usual Bohmian form, with a wave function guiding it in the usual way. If that were the case, \shelly{quantum mechanics,} while entirely absent \shelly{from} the fundamental (shape space) level, would arise as a choice of gauge in  absolute configuration space.
If this were the case, we would have another instance, maybe the most surprising, of what we have called \shelly{a} Kantian component of physical laws---that 
much of what we \shelly{take} as fundamental  is not fundamental, but imposed by us by a proper choice of gauge. 

In view of this---which at the moment is very speculative---and in view
of what has been  said before about relativity---much less speculative---it seems to us not unreasonable to entertain the \shelly{following possibility: that}
quantum mechanics
and relativity, the two great revolutions of  \shelly{twentieth century physics, may not be fundamental at all  but  imposed  on nature by us.}

\end{document}